\documentstyle[aps,prb,multicol,psfig]{revtex}
\newcommand{\be}{\begin{equation}}
\newcommand{\ee}{\end{equation}}
\newcommand{\rf}[1]{(\ref{#1})}

\begin{document}
\draft

%%%%%%%%%%%%%%%%%%%%%%%%%%%%%%%%%%%%%%%%%%%%%%%%%%%%%%%%%%%%%%%%%%%%
%%%%%%%%%%%%%%%%%%%%%%% Title %%%%%%%%%%%%%%%%%%%%%%%%%%%%%%%%%%%%%%
%%%%%%%%%%%%%%%%%%%%%%%%%%%%%%%%%%%%%%%%%%%%%%%%%%%%%%%%%%%%%%%%%%%%
\title{Origin of Spin Incommensurability \\
in Hole-doped S=1 $\rm Y_{2-x}Ca_x Ba Ni O_5$ Chains}

%%%%%%%%%%%%%%%%%%%%%%%%%%%%%%%%%%%%%%%%%%%%%%%%%%%%%%%%%%%%%%%%%%%%
%%%%%%%%%%%%%%%%%%%%%%% Author(s) %%%%%%%%%%%%%%%%%%%%%%%%%%%%%%%%%%
%%%%%%%%%%%%%%%%%%%%%%%%%%%%%%%%%%%%%%%%%%%%%%%%%%%%%%%%%%%%%%%%%%%%
\author{Andr\'e Luiz Malvezzi}
\address{Departamento de F\'\i sica - Faculdade de Ci\^encias -
  Universidade Estadual Paulista \\ Caixa Postal 473, 17.033-360,
  Bauru, SP, Brazil}

\author{Elbio Dagotto}
\address{Department of Physics and National High Magnetic Field
  Laboratory, Florida State University, Tallahassee, FL 32306}

\date{\today}

\maketitle

%%%%%%%%%%%%%%%%%%%%%%%%%%%%%%%%%%%%%%%%%%%%%%%%%%%%%%%%%%%%%%%%%%%%
%%%%%%%%%%%%%%%%%%%%%%%%% Abstract %%%%%%%%%%%%%%%%%%%%%%%%%%%%%%%%%
%%%%%%%%%%%%%%%%%%%%%%%%%%%%%%%%%%%%%%%%%%%%%%%%%%%%%%%%%%%%%%%%%%%%
\begin{abstract}
Spin incommensurability (IC) has been recently experimentally 
discovered in the hole-doped Ni-oxide chain compound
$\rm Y_{2-x}Ca_x Ba Ni O_5$ (G. Xu {\it et al.}, Science {\bf 289}, 419 (2000)).
Here a two orbital model for this material is studied using
computational techniques. Spin IC is observed in a wide range of
densities and couplings. The phenomenon originates in 
antiferromagnetic correlations 
``across holes'' dynamically generated to improve hole movement,
as it occurs in the one-dimensional Hubbard model and
in recent studies of the two-dimensional extended t-J model.
The close proximity of ferromagnetic and phase-separated states
in parameter space are also discussed.
\end{abstract}

%%%%%%%%%%%%%%%%%%%%% PACS number %%%%%%%%%%%%%%%%%%%%%%%%%%%%%%%%%%

%%%%%%%%%%%%%%%%%%%%%%%%%%%%%%%%%%%%%%%%%%%%%%%%%%%%%%%%%%%%%%%%%%%%
%%%%%%%%%%%%%%%%%%%%% main-body begins here %%%%%%%%%%%%%%%%%%%%%%%%
%%%%%%%%%%%%%%%%%%%%%%%%%%%%%%%%%%%%%%%%%%%%%%%%%%%%%%%%%%%%%%%%%%%%
\begin{multicols}{2}
\narrowtext

One of the most remarkable results 
in hole-doped two-dimensional (2D)
high temperature superconductors is the
existence in neutron scattering experiments of 
spin incommensurability (IC).\cite{IC}
This result is compatible with the formation of stripes, 
with a $\pi$-shift in the antiferromagnetic (AF) order across the
stripes. Considerable theoretical work has
been devoted to the explanation of these structures,
but no consensus has been reached. 
Recently, a phenomenon analogous to the spin IC in cuprates has
been reported in the one-dimensional (1D) 
compound $\rm Y_{2-x}Ca_x Ba Ni O_5$.\cite{xu} 
The parent material (x=0) is a spin-one (S=1) chain 
described by a Heisenberg model, with a spin liquid ground-state. Upon
doping, the resistivity is drastically reduced,\cite{ditusa}
and the magnetic IC arises.\cite{xu} 
These novel results raise the interesting possibility of a
common origin of the spin IC phenomena observed in doped 2D cuprates 
and 1D nickelates.

The purpose of this paper is to present theoretical calculations
searching for spin IC in models for doped S=1 chains.
Theoretical studies of these systems
have been presented before,\cite{others,dagotto} but the
magnetic order upon doping has not been investigated in detail.
The main result of our effort is that spin IC
 can indeed be observed when holes are doped into a spin-gapped S=1
 chain. The effect originates in the
 local existence of dynamically induced
 robust AF correlations between the spins
located at both sides of a mobile hole (``across-the-hole''). 
This spin structure is generated to facilitate the
hole movement, namely to improve the kinetic energy portion of the
Hamiltonian. This is qualitatively similar to results
reported in 1D Hubbard
models with spin-charge separation,\cite{shiba}
and also in studies by Martins et al. on ladders and 2D
models with very mobile holes.\cite{martins}
Based on the results reported here, it is natural
to conjecture that the spin IC in 1D Ni-oxides
may originate in similar tendencies of these systems toward
spin-charge separation. In the 1D S=1/2 system this separation is complete,
while in 2D S=1/2 and 1D S=1 spin-gapped systems
it appears to occur only at short distances.
However, this is sufficient to induce a robust spin IC.

The proposed Hamiltonian for doped S=1 chains is
\begin{eqnarray}
\label{H}
H &=& -\sum_{\langle i j \rangle a b \sigma} t_{a b} (
 c^{\dagger}_{i a \sigma} c_{j b \sigma} + {\mbox h.c.} ) 
+ U\sum_{i a} n_{i a \uparrow} n_{i a \downarrow} \nonumber \\ 
& &
+ U'\sum_{i \sigma \sigma '} n_{i 1 \sigma} n_{i 2 \sigma '}
- J\sum_{i \sigma \sigma '} 
c^{\dagger}_{i 1 \sigma}c_{i 1 \sigma '}
c^{\dagger}_{i 2 \sigma '}c_{i 2 \sigma},
\end{eqnarray}
where the notation is standard, namely  $n_{i a \sigma}$ is the number
operator at site $i$, orbital $a$,
and spin $\sigma$, $a = 1 (2)$ corresponds to orbital
$3d(3z^2-r^2)$ ($3d(x^2 - y^2)$) of the $\rm Ni^{+2}$ ion, 
and the chain direction is taken to be the {\em z}-axis.
The hopping amplitudes  between nearest-neighbor
sites are $t_{1 1} = 4/3$, $t_{2 2} = t_{1 2} =
t_{2 1} = 0$.\cite{hotta} 
$U$ and $U'$ are onsite intra-orbital and inter-orbital
Coulomb interactions, respectively, while $J (> 0)$ is the Hund coupling which
favors parallel spins on the same site.
Note that
orbital rotational invariance requires $U$=$U' + J$.\cite{Kuei,newreview} 
For simplicity, in this first attempt to understand the 
spin IC effect
the oxygen ions are not explicitly included in the Hamiltonian. The use of
Zhang-Rice ``doublets'' (instead of singlets) in previous literature of 
Ni-oxides\cite{dagotto,riera} justifies in part this approximation.

The many-body technique used here to analyze ground-state properties
of Hamiltonian \rf{H} is the Density-Matrix Renormalization
 Group (DMRG) method.\cite{DMRG}
The finite-system variation of DMRG was applied, working with
open boundary conditions (OBC). Truncation errors were kept around
$10^{-5}$ or smaller, using about 100 states. 
In order to characterize ground-state properties,
a variety of observables have been measured, such as the
spin structure-factor defined as 
\begin{equation}
S(k) = \frac{1}{L} \sum_{j m a b}
\langle {\bf S}_{j a} \cdot {\bf S}_{m b} \rangle \: e^{i(j-m)k},
\end{equation}
where ${\bf S}_{j a} = \sum_{\alpha \beta} c^{\dagger}_{j a \alpha}
{\bf \sigma}_{\alpha \beta} c_{j a \beta}$,
$\langle \: \rangle$ denotes ground-state expectation value, and $L$ is the
length of the chain.
The charge structure-factor was also measured, but no evidence of charge
ordering has been observed.
Also the local
density, $\langle n_{i}\rangle = \sum_{a \sigma} 
\langle n_{i a \sigma} \rangle $,
and total-spin $z$-component, $\langle {\bf S}_{j}^{z} 
\rangle =
\sum_{a} \langle {\bf S}_{j a}^{z} \rangle $, were studied,
but the main results were obtained focusing on spin correlations.
\begin{figure}[h]
\vskip-0.45truein
\centerline{\psfig{figure=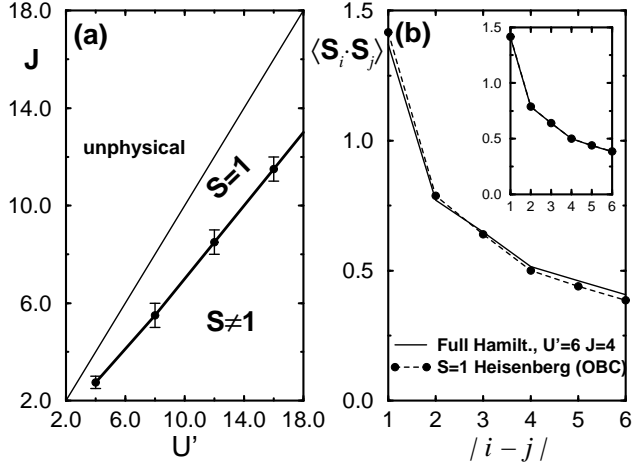,width=8.5cm,angle=-90}}
\vskip0.2truein
\caption{ 
Results at $n$=$2$ for a 16-site chain.
(a) Phase diagram.
(b) Spin correlations at $(U',J)$=$(6,4)$.
For comparison, the correlations of the AF S=1 Heisenberg chain
are also shown. The inset shows the same but for
$(U',J)$=$(16,13)$. Details can be found in the text.
}
\end{figure}

Let us consider density $n$=$2$ first, expected to
represent the undoped material. Fig.~1a shows the $(U',J)$
phase diagram for a system of 16 sites and 32 electrons.  
Each of the three marked regions have distinct features. 
In the regime above line $J$=$U'$ the
Hund coupling $J$ provides an electronic attractive interaction 
which competes with the $U'$ repulsion. 
In order to have the expected overall repulsive 
total inter-orbital interaction, $J < U'$ is required.\cite{Kuei}
In addition, to create a spin-1
chain, as observed in experiments,
a robust Hund coupling is needed.
In fact, inside the region labeled {\bf S=1} it was observed
that the total spin at site $j$,
defined as ${\bf S}_{j} = {\bf S}_{j 1} + {\bf S}_{j 2}$,
is to an excellent approximation equal to one\cite{S=1} 
for any site $j$. This is the region of interest in the present
problem. In the regime
labeled {\bf S$\ne$1}, the mean spin is less than 1 since $J$
is not sufficiently strong.\cite{comment}
The  frontier {\bf S=1}--{\bf S$\ne$1} has error bars caused
by finite size effects, estimated from 
chains with L=8, 16 and 20 sites. 

In the region of local {\bf S=1} states it is necessary to
verify whether a Heisenberg spin-1 chain provides a proper 
effective model for the system. For this purpose,
ground-state spin correlations
 were calculated for several couplings
inside the {\bf S=1} region. As shown in Fig.~1b, these correlations behave
quite similarly  for two typical sets of couplings $(U',J)$, and in
addition they are nearly identical to those arising from the much simpler
spin-1 Heisenberg chain model.
In both cases, the results at a given
distance were averaged over the entire chain to improve the
convergence to the thermodynamic
limit when different OBC system sizes are used.
\begin{figure}[h]
\vskip-0.45truein
\centerline{\psfig{figure=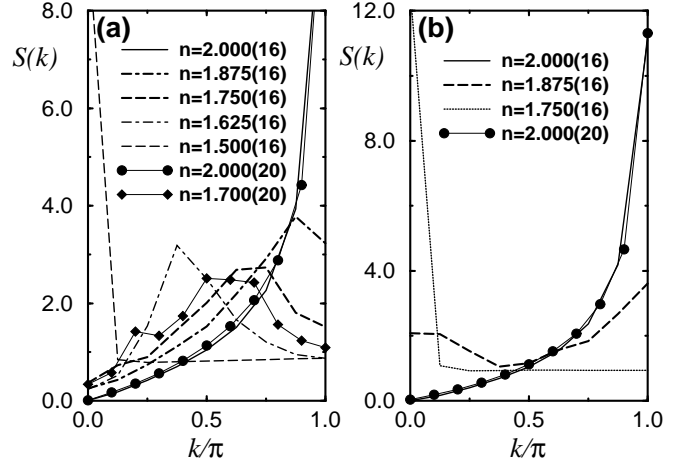,width=8.5cm,angle=-90}}
\vskip0.2truein
\caption{The spin structure factor for systems of 16 and 20 sites (indicated
in parenthesis) and different electron densities:
(a) $U'$=$6$ and $J$=$4$; (b) $U'$=$16$ and $J$=$13$.
}
\end{figure}

Inside the {\bf S=1} region, it can be shown that not only 
the ground-state properties, but also the low-energy
physics of Eq.(1) reduces to the spin-1 Heisenberg chain,
with an effective antiferromagnetic
coupling $J_{Heis}$ as an overall scale, which depends
on $U'$ and $J$. This interesting result confirms that our
model at n=2 is realistic. 
In fact,  for several values of $(U',J)$ and several
 lattice sizes, the effective Heisenberg coupling $J_{Heis}$ was estimated 
  by comparing energy gaps of Eq.(1) 
 with those of the spin-1 Heisenberg chain. 
For given values of $(U',J)$, data coming from 
systems with sizes between 4 and 20 sites  
yield approximately the same value for $J_{Heis}$.
Even though a closed form for the dependence of $J_{Heis}$ 
with couplings is unknown, it is clear from our results that $J_{Heis}$ is
inversely proportional to $U'$ and $J$. Typical values
of $J_{Heis}$ are $0.14 \pm 0.01$ for $(U',J)$=$(4,3)$,
$0.042 \pm 0.002$ for $(U',J)$=$(16,13)$, and
$0.019 \pm 0.002$ for $(U',J)$=$(32,28)$, showing that the expected spin-gaps
are very small in the units that are natural for Hamiltonian Eq.(1).
Reducing $J$ at fixed $U'$, the 
{\bf S$\ne$1} region is accessed where
the on-site S=1 and S=0 states start mixing,
producing deviations in the spin correlations 
from those of the Heisenberg model.

The introduction of mobile holes in the {\bf S=1} region,
through the reduction of the electronic density, produces 
the most interesting results of our investigations. 
In the regime of moderate couplings, both $U'$ and $J$ less than 10, 
 the presence of {\it spin incommensurate} tendencies are clearly observed 
(Fig.~2a), in a wide range of densities, through the appearance 
of a prominent peak at a momentum different from $\pi$. No fine tuning
of parameters is needed, the effect exists in a wide range of 
couplings.\cite{height} 
Note that although experiments\cite{ditusa} suggest that the
material $\rm Y_{2-x}Ca_x Ba Ni O_5$ remains insulating upon doping, the
observed drastic change in the resistivity\cite{ditusa} justifies the use of a
metallic description of the compound, as observed in our studies where
charge is not localized.\cite{metal} It may occur that holes
in the real material are
mobile over several lattice spacings, but defects
destroy the complete metallicity of the samples (see also Ito {\it et
al.}\cite{ito}).
\begin{figure}[h]
\vskip-0.1truein
\centerline{\psfig{figure=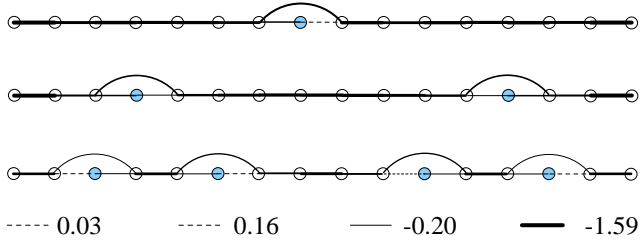,width=8.5cm,angle=-90}}
\vskip0.2truein
\caption{Spin correlations for one ($n$=$1.9375$), two ($n$=$1.875$), 
and four ($n$=$1.75$) holes, doped on a 16 sites system at $n$=$2$.
Grey circles indicate the sites where the holes were projected
at orbital 1. The thickness of the lines is 
proportional to the spin bond strengths,
according to the scale shown. Full (dashed) lines indicate AF (FM) bonds.  
The couplings are $(U',J)$=$(4,3)$. In all cases shown,
across-the-hole bonds are AF
and $S(k)$ has spin IC, as in Fig.2a.}
\end{figure}

Insight into the origin of the spin IC
can be gained by investigating spin correlations 
in the vicinity of the holes.  
This can be obtained by projecting the hole on a giving site out of the
ground-state, and then
measuring spin correlations around and across that hole. This projection
procedure has been successfully previously applied in studies of 
models for the cuprates.\cite{martins,projection}
Fig.~3 shows a pictorial representation of a 16 sites system
doped with one, two, and four holes.\cite{asym} The holes were projected
onto their most probable locations on the chain (remember that 
OBC are used). Since 
hopping is restricted within the  $3d(3z^2-r^2)$ orbitals, the
projection is always carried out onto orbital 1. After the projection,
the spin bond strengths $\langle {\bf S}_i \cdot {\bf S}_j \rangle$
were measured for $i$ and $j$ being nearest-neighbor sites, or
sites adjacent to the same hole.   
The bond strength is proportional to the thickness of the 
line connecting the sites in Fig.~3. In that figure, the AF
$\langle {\bf S}_i \cdot {\bf S}_j \rangle < 0$ (FM 
$\langle {\bf S}_i \cdot {\bf S}_j \rangle > 0$) 
bonds are represented by a full (dashed) line. 

The results in Fig.~3 were obtained for $(U',J)$=$(4,3)$,
but they are representative of tendencies found in most cases studied
here, for small and moderate couplings. 
Across the projected holes, the clear presence of
AF bonds is observed. It is this $\pi$-shift in the staggered
spin pattern that causes the incommensurability. 
Intuitively these $\pi$-shifts improve the mobility of
holes. Without them, the hole movement would lead to the creation of
``strings'' of ferromagnetic bonds, increasing the energy.
This across-the-hole structure is reminiscent
of those observed in models for cuprates 
in 1D and 2D,\cite{shiba,martins} with the
conceptual difference that the present results exist in the spin-liquid
regime (nonzero spin gap). It appears  that the local tendency to produce
spin IC, in the form of across-the-hole AF correlations, is present regardless
of the spin-liquid vs critical or long-range ordered character of the 
spin background. 
Note that in fact for a S=1 chain 
it is possible to use the concept of ``AF order'' in the spin 
background  as  long as the distances considered are smaller than
the AF correlation length, which for the S=1 Heisenberg chain is about
six lattice spacings. This is sufficiently large to accommodate the
small across-the-hole structure observed in our studies. This also
suggests that even if holes are localized in $\rm Y_{2-x}Ca_x Ba Ni O_5$,
with a distorted region of only a few lattice spacings,\cite{xu,ito} it
is still possible for holes to generate spin IC.

The results shown in Fig.~2a are representative of data gathered at 
several couplings, and they  not only contain information on 
spin IC, but on other nontrivial spin patterns as well. In particular,
it was observed that for a fixed $U'$, the spin IC
survives longer the decrease in electronic density
by reducing $J$, but eventually a {\it ferromagnetic} (FM) regime
always appears if the hole doping is large enough, as shown in Fig.~2a
where a prominent peak at zero momentum is observed at density 1.5.
Then, the phase diagram of models for doped S=1 chains must contain
a FM phase at low enough electronic density. The existence of a
competing FM phase
is in agreement with previous calculations using effective t-J-like
models valid at low energies,\cite{riera} also with expectations coming from
manganite investigations where the relevant densities are close 
to 1,\cite{hotta,yunoki} and it could be tested
experimentally by increasing the hole doping in the S=1 chain compounds.

In addition,
doping at stronger couplings than those shown in Fig.~2a  
produces {\it phase separation} (PS) 
between hole-rich and hole-poor regions, if $J$ is
 close to the lower boundary of the {\bf S=1} region of Fig.~1a.
 This is the case for the results shown in Fig.~2b. 
 Such PS regime is characterized by a negative inverse compressibility,
 the presence of two simultaneous peaks in the spin structure factor at
a fixed total electronic density, and
 a tendency to charge clustering in the local density. 
Tendencies toward FM and
electronic PS found here are similar to those  encountered in Kondo-like
models for the manganites,\cite{yunoki} and effective models for 
nickelates.\cite{riera} Manganite models also have two orbitals,
but in addition they contain localized $t_{2g}$ spins. In the S=1 chains
considered here, the
orbital $(x^2-y^2)$ is fully occupied, and plays a role analog to that of
those localized spins of Mn-oxides. Then, 
the tendencies to ferromagnetism observed
here may be caused by a mechanism similar to Double Exchange, which
could be operative in highly-doped Ni-oxides.

The analysis of the previous paragraphs 
leads to a second possible rationalization of the spin IC.
As explained before,
hole doping eventually drives the system into a FM phase. 
But the transition between the IC phase discussed previously
and the FM phase is not always abrupt. Fig.~4 contains results 
analogous to those of Fig.~3 but now for $(U',J)$=$(8,6)$. 
In all three doped systems discussed, 
$S(k)$ was still found to be incommensurate.\cite{comment2}
In Fig.~4, the spin correlation pattern when one or two holes
are projected to their most likely position is similar to
Fig.~3, but for the case of four holes 
the spin configuration is clearly different. 
Instead of AF bonds across the holes, a
small three-site island of ferromagnetism
is formed around each hole, resembling
a magnetic polaron. 
The presence of FM polarons was already observed in recent studies of
doped two-orbital chains with strong Hund's rule couplings.\cite{ammon}
These polarons are coupled in such a way that they
generate an overall spin IC (in particular
note the way in which the two central polarons are coupled
antiferromagnetically).
Then, there is a second mechanism to generate spin IC in this context. However,
since experimentally no evidence of FM has been
reported,\cite{xu} the first explanation with AF
correlations across holes appears more realistic.

\begin{figure}[h]
\vskip-0.1truein
\centerline{\psfig{figure=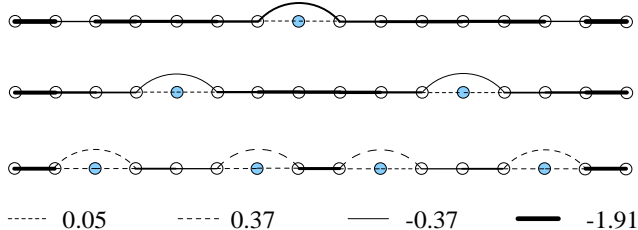,width=8.5cm,angle=-90}}
\vskip0.2truein
\caption{
Same as Fig.~3 but for $(U',J)$=$(8,6)$. Although in
all cases shown $S(k)$ is still incommensurate, the system
doped with four holes resembles a polaronic lattice, with
the holes located at the centers of small FM islands. 
}
\end{figure}

Note that the presence of phase-separation tendencies reveal a source of
attraction among holes. Actually, in previous literature\cite{riera} 
it was already shown that holes attract in a large region of
parameter space, using t-J-like models for Ni-chains. This is
reasonable, since each hole creates a spin distortion, and sharing distortions
reduces the energy. The pairing of holes in a spin-gapped system has
before been documented in models for ladders,\cite{science} and it appears
in some models for doped S=1 chains.\cite{riera} 
Studies of two-orbital models have
also revealed robust singlet-pair correlations.\cite{ammon} Then, it is
tempting to speculate that the material $\rm Y_{2-x}Ca_x Ba Ni O_5$ may
become superconducting upon the application of pressure, as it happens
with the two-leg ladder compound [14-24-41]. Experimental 
studies of doped S=1 chains under high pressure 
may lead to interesting results.

Summarizing, studies of hole-doped S=1 chains show that spin
incommensurate correlations can be generated in a 
realistic model,\cite{s1model} 
as observed in experiments. The mechanism is similar
to the phenomenon recently discussed in S=1/2 ladders and planes, namely
the generation of AF correlations across holes to optimize the hole
movement.\cite{martins} 
It is expected that this mechanism may contribute, at least in
part,\cite{lastremark,bose} 
to the understanding of the recently unveiled spin IC tendencies in
$\rm Y_{2-x}Ca_x Ba Ni O_5$. 

%%%%%%%%%%%%%%%%   Acknowledgments  %%%%%%%%%%%%%%%%%%%%%%%%%
%

The authors especially thank G. Xu and C. Broholm for useful
discussions. 
A.L.M. acknowledges the financial support from Funda\c c\~ao
de Amparo \`a Pesquisa do Estado de 
S\~ao Paulo (FAPESP-Brazil). E.D. is supported by grant NSF-DMR-9814350.

%%%%%%%%%%%%%%%%%%%%%%%%%%%%%%%%%%%%%%%%%%%%%%%%%%%%%%%%%%%%%%%%%%%%
%%%%%%%%%%%%%%%%%%%%%%%%% References %%%%%%%%%%%%%%%%%%%%%%%%%%%%%%%
%%%%%%%%%%%%%%%%%%%%%%%%%%%%%%%%%%%%%%%%%%%%%%%%%%%%%%%%%%%%%%%%%%%%

\end{multicols}

\begin{references}
\bibitem{metal}
Future calculations of the optical conductivity using Eq.(1)
or the model of Refs.6,13 will clarify how robust is the Drude
contribution at low energy.
\bibitem{asym} DMRG calculations for one hole on an odd number
of sites chain lead to results very similar to those at the top
of Fig.3, where an even number of sites was used. In both cases
robust AF correlations across the hole are clearly observed.
\bibitem{s1model}
Studies in simpler models of S=1 spins and S=1/2 holes
as proposed in Refs.6,13 may also lead to spin IC features.
Results in Ref.6 already suggest such behavior.
\bibitem{comment}
Due to the use of a realistic model with Coulomb interactions and
$U$=$U'$+$J$, here it is not possible to simply select $J$=$\infty$ 
to enforce the S=1 state at every non-doped site.
\bibitem{height} Note that the height of incommesurate peak in
Fig.2a increases when the density is reduced from n=1.750 to
1.625. This effect is caused by the presence of ferromagnetic islands
in the ground state, as illustrated in Fig.4.
\bibitem{comment2} While a system doped with six holes at $(U',J)$=$(6,4)$
is also incommensurate (Fig.~2a), for $(U',J)$=$(8,6)$ it
is already FM.
\end{references}
\end{document}